\RequirePackage{fix-cm}
\documentclass[twocolumn]{svjour3}          
\smartqed  
\usepackage{graphicx}
\usepackage{hyperref}
\usepackage{epstopdf}
\usepackage{bm} 
\usepackage{mathrsfs}
\usepackage{amsmath}
\usepackage{cancel}
\usepackage{amssymb}
\begin{document}

\title{A New Set of Maxwell-Lorentz Equations and Rediscovery of Heaviside-Maxwellian (Vector) Gravity from Quantum Field Theory
}


\author{Harihar Behera         \and
        Niranjan Barik 
}


\institute{H. Behera \at
              BIET Degree College and BIET Higher Secondary School, Dhenkanal-759001, Odisha, India \\
              \email{behera.harihar@gmail.com}           
           \and
           N. Barik \at
              Department of Physics, Utkal University, Vani Vihar, Bhubaneswar-751004, Odisha, India \\
 \email{dr.nbarik@gmail.com}             
}

\date{Received: date / Accepted: date}

\maketitle

\begin{abstract}
We show that if we start with the free Dirac Lagrangian, and demand local phase invariance, assuming the total phase coming from two independent contributions associated with the charge and mass degrees of freedom of charged Dirac particles, then we are forced to introduce two massless independent vector fields for charged Dirac particles that generate all of electrodynamics and gravitodynamics of Heaviside's Gravity of 1893 or Maxwellian Gravity and specify the charge and mass currents produced by charged Dirac particles. From this approach we found: (1) a new set of Maxwell-Lorentz equations, (2) two equivalent sets of gravito-Maxwell-Lorentz equations (3) a gravitational correction to the standard Lagrangian of electrodynamics, which, for a neutral massive Dirac particle, reduces to the Lagrangian for gravitodynamics, (4) attractive interaction between two static like masses, contrary to the prevalent view of many field theorists and (5) gravitational waves emanating from the collapsing process of self gravitating systems carry positive energy and momentum in the spirit of Maxwell's electromagnetic theory despite the fact that the intrinsic energy of static gravitoelectromagnetic fields is negative as dictated by Newton's gravitational law and its time-dependent extensions to Heaviside-Maxwellian Gravity (HMG). Fundamental conceptual issues in linearized Einstein's Gravity are also discussed.
\end{abstract}

\section{Introduction}
\label{intro}
Many field theorists, like Gupta \cite{1}, Feynman \cite{2}, Low \cite{3}, Padmanabhan \cite{4},  Zee \cite{5} and Gasperini \cite{6} and Straumann \cite{7} have rejected spin-1 vector theory of gravity on the ground that if gravitation is described as a spin-1 theory like Maxwell's electromagnetic theory, then two static masses of same sign will repel each orther analogous the case in electromagnetism where two static charges of same sign repel each other, while according to Newton's gravitational theory - two static masses of same sign attract each other. However, here we show that this not true, if one considers appropriate field equations of vector gravity derived here in a novel application of the well establisshed principle of local phase (or gauge) invariance of field theory to massive Dirac fields. Subscribing to Feynman's view \cite{2} that ``space-time curvature is not essential to physics", and adopting Minkoskian space-time here we show that if we start with the free Dirac Lagrangian, and demand local phase invariance, considering the total phase coming from two independent contributions associated with the charge and mass degrees of freedom of charged Dirac particles, then we are forced to introduce two massless independent vector fields for charged Dirac particles that generate  all of electromagnetism and gravielectromagnetism of Heaviside's Gravity (HG)\footnote{Heaviside had speculated a gravitational analogue of Lorentz force law with a sign error that is corrected in this work.}\cite{8,9,10,11,12,13,14,15} of 1893 or Maxwellian Gravity(MG)\footnote{Which looks mathematically different from Heaviside's Gravity due to some differences in the sign of certain terms. But HG and MG are shown here to represent a single physical theory called Heaviside-Maxwellian Gravity (HMG) by correct representations of their respective field and force equations.}\cite{16} and specify the charge and mass currents produced by charged Dirac particles. Our new approach naturally renders a gravitodynamics correction to the standard Lagrangian of quantum electrodynamics, which, for a neutral massive Dirac particle, reduces to the Lagrangian of quantum gravitodynamics. The resulting spin-1 vector gravity is shown to produce attractive interaction between two static like masses, contrary to the prevalent view. In the present approach, we also found a new set of Maxwell-Lorentz equations (n-MLEs) of electrodynamics physically equivalent to the standard Maxwell-Lorentz equations (s-MLEs). The n-MLEs and s-MLEs are listed in Table 1 for comparison. Similarly, 
our present findings of the gravitational Maxwell-Lorentz equations (g-MLEs) of HG and MG along with n-MLEs are listed in the Table-2, which exactly match with the recent results obtained by Behera \cite{17} following Schwinger's inference of s-MLEs within Galileo-Newtonian physics, if the speed of gravitational waves in vacuum $c_g = c$, the speed of light in vacuum.
\begin{table}[!h]
\caption{Standard Maxwell-Lorentz Equations (s-MLEs) and new Maxwell-Lorentz Equations (n-MLEs) in SI units. }
\label{Table-1}
\begin{tabular}{|l|l|}  
\hline
s-MLEs & n-MLEs     \\
\hline
$\mathbf{\nabla}\cdot\mathbf{E} = \rho_e/\epsilon_{0} $ \quad & $\mathbf{\nabla}\cdot\mathbf{E}\, = \rho_e/\epsilon_{0}$     \\
\hline
$\mathbf{\nabla}\cdot\mathbf{B} = 0$ & $\mathbf{\nabla}\cdot\mathbf{B} = 0$      \\ \hline
 $\mathbf{\nabla}\times\mathbf{B}\,=\,+\,\mu_{0}\mathbf{j}_e\,-\,\frac{1}{c^2}\frac{\partial \mathbf{E}}{\partial t}$ &  $\mathbf{\nabla}\times\mathbf{B}\,=\,-\,\mu_0\mathbf{j}_e\,-\,\frac{1}{c^2}\frac{\partial \mathbf{E}}{\partial t}$   \\ \hline 
    $\mathbf{\nabla}\times\mathbf{E}\,=\,-\,\frac{\partial \mathbf{B}}{\partial t}$ &   $\mathbf{\nabla}\times\mathbf{E}\,=\,+\,\frac{\partial \mathbf{B}}{\partial t}$     \\ \hline
  $\frac{d\mathbf{p}}{dt}\,=\,q\left[\mathbf{E}\,+\,\mathbf{u}\times \mathbf{B}\right] $   &   $\frac{d\mathbf{p}}{dt}\,=\,q\left[\mathbf{E}\,-\,\mathbf{u}\times \mathbf{B}\right]$        \\ \hline 
 $\mathbf{B}\,=\,+\,\mathbf{\nabla}\times \mathbf{A}_{e}$   &  $\mathbf{B}\,=\,-\mathbf{\nabla}\times \mathbf{A}_e$  \\ \hline
  $\mathbf{E}\,=\,-\,\mathbf{\nabla}\phi_{e} \,-\,\frac{\partial \mathbf{A}_e}{\partial t}$  &  $\mathbf{E}\,=\,-\,\mathbf{\nabla}\phi_{e} \,-\,\frac{\partial \mathbf{A}_e}{\partial t}$       \\ \hline 
\end{tabular}
\vspace*{-4pt}
\end{table}
\begin{table}[!h]
\caption{Gravito-Maxwell-Lorentz Equations (g-MLEs) of Heaviside Gravity (HG) and Maxwellian Gravity (MG), where $\mu_{0g} = 4\pi G/c^2$. }
\label{Table-2}
\begin{tabular}{|l|l|}
\hline
g-MLEs of HG & g-MLEs of MG    \\
\hline
$\mathbf{\nabla}\cdot\mathbf{g} = - 4\pi G\rho_0 = - \rho_0/\epsilon_{0g} $ \quad & $\mathbf{\nabla}\cdot\mathbf{g}\, = \,-\,4\pi G\rho_0 = - \rho_0/\epsilon_{0g}$       \\
\hline
$\mathbf{\nabla}\cdot\mathbf{b} = 0$ & $\mathbf{\nabla}\cdot\mathbf{b} = 0$     \\ \hline
    $\mathbf{\nabla}\times\mathbf{b}\,=\,+\,\mu_{0g}\mathbf{j}_g\,-\,\frac{1}{c^2}\frac{\partial \mathbf{g}}{\partial t}$ & $\mathbf{\nabla}\times\mathbf{b}\,=\,-\,\mu_{0g}\mathbf{j}_g\,+\,\frac{1}{c^2}\frac{\partial \mathbf{g}}{\partial t}$    \\ \hline 
    $\mathbf{\nabla}\times\mathbf{g}\,=\,+\,\frac{\partial \mathbf{b}}{\partial t}$ & $\mathbf{\nabla}\times\mathbf{g}\,=\,-\,\frac{\partial \mathbf{b}}{\partial t}$     \\ \hline
  $\frac{d\mathbf{p}}{dt}\,=\,m_0\left[\mathbf{g}\,-\,\mathbf{u}\times \mathbf{b}\right] $   & $\frac{d\mathbf{p}}{dt}\,=\,m_0\left[\mathbf{g}\,+\,\mathbf{u}\times \mathbf{b}\right]$          \\ \hline 
   $\mathbf{b}\,=\,-\,\mathbf{\nabla}\times \mathbf{A}_{g}$ & $\mathbf{b}\,=\,+\,\mathbf{\nabla}\times \mathbf{A}_{g}$    \\ \hline
     $\mathbf{g}\,=\,-\,\mathbf{\nabla}\phi_{g} \,-\,\frac{\partial \mathbf{A}_{g}}{\partial t}$ & $\mathbf{g}\,=\,-\,\mathbf{\nabla}\phi_{g} \,-\,\frac{\partial \mathbf{A}_g}{\partial t}$        \\ \hline 
\end{tabular}
\vspace*{-4pt}
\end{table}

\textit{Units and Notations:} Here we use SI units so that the paper can easily be understood by general readers. The flat space-time symmetric metric tensor $\eta_{\alpha \beta} = \eta^{\alpha \beta}$ is a diagonal matrix with diagonal elements $
\eta_{00} = 1,\, \eta_{11} = \eta_{22} = \eta_{33} = -1 $, space-time 4-vector $x = x^\alpha=(ct, \mathbf{x})$ and $x_\alpha=(ct, -\mathbf{x})$, 4-velocity $ dx^\alpha/d\tau = \dot{x}^\alpha = (c\gamma_u,\,\mathbf{u}\gamma_u)$ is the 4-velocity with $\gamma_u = (1 - u^2/c^2)^{-1/2}$,\, and $\tau$ is the proper time along the particle's world-line, energy momentum four vector $p^\alpha = (p_0,\,\mathbf{p}) = (E/c,\,\mathbf{p})$,\,
$\partial_\alpha \equiv \left(\partial/{c\partial t},\,\mathbf{\nabla}\right), \, \partial^\alpha \equiv \left(\partial/{c\partial t},\,-\mathbf{\nabla}\right)$,\,the {\it D'Alembertian} operator is $
\Box\,=\,\partial_\alpha \partial ^\alpha \,=\,\partial^2/{c^2\partial t^2}\,-\,\mathbf{\nabla}^2$, where  Einstein's convention of sum over repeated indices is used. 
\section{Consequences of Local Phase Invariance for Charge and Mass Degrees of Freedom }
It is well known that the free Dirac Lagrangian density for a Dirac particle of rest-mass $m_0$
\begin{equation}\label{eq1}
\mathcal{L} = i\hbar c\overline{\psi}\gamma^\mu\partial_\mu \psi - m_0c^2\overline{\psi}\psi 
\end{equation}
is invariant under the transformation 
\begin{equation}\label{eq2}
\psi \rightarrow e^{i\theta}\psi \qquad {(\mbox{global phase transformation})}
\end{equation}
where $\theta$ is any real number. This is because under global phase transformation eq. (\ref{eq2}), $\overline{\psi} \rightarrow e^{-i\theta}\overline{\psi}$ which leaves $\overline{\psi}\psi$ in (\ref{eq1}) unchanged as the exponential factors cancel out. But eq. (\ref{eq1}) is not invariant under the following transformation 
\begin{equation}\label{eq3}
\psi \rightarrow e^{i\theta(x)}\psi \qquad {(\mbox{local phase transformation})}
\end{equation} 
where $\theta$ is now a function of space-time $x (= x^\mu)$, because the factor $\partial_\mu \psi$ in (\ref{eq1}) now picks up an extra term from the derivative of $\theta(x)$: 
\begin{equation}\label{eq4}
\partial_\mu \psi \rightarrow \partial_\mu\left(e^{i\theta(x)}\psi\right)= i\left(\partial_\mu\theta\right)e^{i\theta}\psi + e^{i\theta}\partial_\mu \psi
\end{equation}
so that under local phase transformation,
\begin{equation}\label{eq5}
\mathcal{L} \rightarrow \mathcal{L}^\prime = \mathcal{L} - \hbar c \left(\partial_\mu\theta\right)\overline{\psi}\gamma^\mu\psi.
\end{equation}
Now suppose that the phase $\theta(x)$ is made up of two parts:
\begin{equation}\label{eq6}
\theta(x) = \theta_1(x) + \theta_2(x),
\end{equation} 
which come from two independent contributions. Then (\ref{eq6}) becomes 
\begin{equation}\label{eq7}
\mathcal{L} \rightarrow \mathcal{L}^\prime = \mathcal{L} - \hbar c \left(\partial_\mu\theta_1\right)\overline{\psi}\gamma^\mu\psi - \hbar c \left(\partial_\mu\theta_2\right)\overline{\psi}\gamma^\mu\psi
\end{equation}
For a charged Dirac particle of charge $q$ and mass $m_0$, we can re-write $\mathcal{L}^\prime$ in eq. (\ref{eq7}) as
\begin{align} 
\mathcal{L}^\prime &= \mathcal{L} -\hbar c \left(\partial_\mu\theta_1\right)\overline{\psi}\gamma^\mu\psi - \hbar c \left(\partial_\mu\theta_2\right)\overline{\psi}\gamma^\mu\psi  \nonumber \\
&= \mathcal{L} + \left[\partial_\mu \left(-\frac{\hbar}{q}\theta_1 \right) q + \partial_\mu \left(-\frac{\hbar}{m_0}\theta_2 \right)m_0 \right] c\overline{\psi}\gamma^\mu\psi  
\notag \\ 
&= \mathcal{L} + j_e^\mu \partial_\mu \lambda_1 (x) + j_g^\mu \partial_\mu \lambda_2 (x) \label{eq8},
\end{align} 
where 
\begin{align} 
 j_e^\mu = qc(\overline{\psi}\gamma^\mu\psi) = \text{4-charge-current density}, \label{eq9}  \\
 j_g^\mu = m_0c(\overline{\psi}\gamma^\mu\psi) = \text{4-mass-current density}, \label{eq10}
\end{align} 
and $\lambda_1(x)$ and $\lambda_2(x)$, respectively stands for 
\begin{equation}\label{eq11}
\lambda_1 (x) = - \frac{\hbar }{q}\theta_1(x),  \quad  \mbox{and} \quad \lambda_2 (x) = - \frac{\hbar }{m_0}\theta_2 (x).
\end{equation}
In terms of $\lambda_1$ and $\lambda_2$ then, under the local phase transformation
\begin{align}
\psi \rightarrow \psi^\prime = e^{-\frac{i}{\hbar}\left[q\lambda_1(x)+ m_0\lambda_2(x)\right]} \psi, \label{eq12}\\
\mathcal{L} \rightarrow \mathcal{L}^\prime = \mathcal{L} + j_e^\mu\partial_\mu \lambda_1 + j_g^\mu\partial_\mu \lambda_2. \label{eq13}
\end{align}
Now, {\it we demand that the complete Lagrangian be invariant under local phase transformations}. Since, the free Dirac Lagrangian density (\ref{eq1}) is not locally phase invariant, we are forced to add something to swallow up or nullify the extra term in eq. (\ref{eq13}). To this end, we suppose 
\begin{equation}\label{eq14}
\mathcal{L} = [i\hbar c\overline{\psi}\gamma^\mu\partial_\mu \psi - m_0c^2\overline{\psi}\psi]\, - \,j_e^\mu A_{e\mu}               \,-\,j_g^\mu A_{g\mu} 
\end{equation} 
where $A_{e\mu}$ and $A_{g\mu}$ are some new fields which interact with the charge and mass current densities and change in coordination with the local phase transformation of $\psi$ according to the rule 
\begin{equation}\label{eq15}
A_{e\mu} \rightarrow A_{e\mu} + \partial_\mu\lambda_1 \quad \mbox{and} \quad  A_{g\mu} \rightarrow A_{g\mu} + \partial_\mu\lambda_2.
\end{equation}
The `new, improved' Lagrangian (\ref{eq14}) is now locally phase invariant. But this was ensured at the cost of introducing two new vector fields that couples to $\psi$ through the last terms in eq. (\ref{eq14}). But the eq. (\ref{eq14}) is devoid of `free' terms for the fields $A_{e\mu}$ and $A_{g\mu}$ (having the dimensions of velocity: $[L][T]^{-1}$). Since these are independent vectors, we look to the Proca-type Lagrangians for these fields \cite{18}:
\begin{align}
\mathcal{L}_e^{\mbox{free}} = \,\frac{\kappa_1}{4}F^{\mu \nu}F_{\mu \nu} + \kappa_{01}\left(\frac{m_1c}{\hbar}\right)^2A_e^\mu A_{e\mu} \label{eq16}  \\
\mathcal{L}_g^{\mbox{free}} = \,\frac{\kappa_2}{4}f^{\mu \nu}f_{\mu \nu} + \kappa_{02}\left(\frac{m_2c}{\hbar}\right)^2A_g^\mu A_{g\mu} \label{eq17}
\end{align}
\noindent
where $\kappa_1, \kappa_2,\kappa_{01}, \text{and}\, \kappa_{02}$ are some dimensional constants to be determined, $m_1$ and $m_2$ are the mass of the free fields $A_{e\mu}$ and $A_{g\mu} $ respectively.
But there is a problem here, for whereas 
\begin{align}
F^{\mu \nu} = (\partial^\mu A_e^\nu - \partial ^\nu A_e^\mu )\,\,\text{or}\,\, F_{\mu \nu} = (\partial_\mu A_{e\nu} - \partial _\nu A_{e\mu})  \label{eq18}    \\
f^{\mu \nu} = (\partial^\mu A_g^\nu - \partial ^\nu A_g^\mu )\,\, \text{or}\,\, f_{\mu \nu} = (\partial_\mu A_{g\nu} - \partial _\nu A_{g\mu})  \label{eq19} 
\end{align}
\noindent
are invariant under the transformation eqs. (\ref{eq15}), $A_e^\mu A_{e\mu}$ and $A_g^\mu A_{g\mu}$ are not.  {\it Evidently, the new fields} $A_e^\mu$ and $A_g^\mu$ {\it must be mass-less} ($m_1 = 0 = m_2$), otherwise the invariance will be lost for these two independent fields. 
The complete Lagrangian density then becomes
\begin{equation}
\mathcal{L} = [i\hbar c\overline{\psi}\gamma^\mu\partial_\mu \psi - m_0c^2\overline{\psi}\psi]\,  
\,+\,\mathcal{L}_e \,+\,\mathcal{L}_g,  \label{eq20}
\end{equation}
where
\begin{eqnarray}
\mathcal{L}_e\,=\,\frac{\kappa_1}{4}F^{\mu \nu}F_{\mu \nu}\,-\,j_e^\mu A_{e\mu}  & \quad \mbox{and} \label{eq21} \\ 
\mathcal{L}_g\,=\,\frac{\kappa_2}{4}f^{\mu \nu}f_{\mu \nu}\,-\,j_g^\mu A_{g\mu} & {} .\label{eq22}
\end{eqnarray}
\noindent
The equations of motion of these new fields can be obtained using the Euler-Lagrange equations:
\begin{equation}\label{eq23}
\partial^\beta\frac{\partial \mathcal{L}_e}{\partial(\partial^\beta A_e^\alpha)}=\frac{\partial \mathcal{L}_e}{\partial A_e^\alpha} \quad \mbox{and} \quad \partial^\beta\frac{\partial \mathcal{L}_g}{\partial(\partial^\beta A_g^\alpha)}=\frac{\partial \mathcal{L}_g}{\partial A_g^\alpha}. 
\end{equation}
A bit calculation (see for example, Jackson \cite{18}) yields 
\begin{eqnarray}
\frac{\partial \mathcal{L}_e }{\partial A_e^\alpha}\,=\,-\,j_{e\alpha} \quad \mbox{and} \quad \frac{\partial \mathcal{L}_g }{\partial A_g^\alpha}\,=\,-\,j_{g\alpha}. \label{eq24} \\
\frac{\partial \mathcal{L}_e}{\partial(\partial^\beta A_e^\alpha)} =\,-\,\kappa_1 F_{\alpha \beta} \quad \mbox{and} \quad \frac{\partial \mathcal{L}_g}{\partial(\partial^\beta A_g^\alpha)} =\,-\,\kappa_2 f_{\alpha \beta}. \label{eq25}
\end{eqnarray}

\noindent
From eqs. (\ref{eq23})-(\ref{eq25}) we get the equations of motion of the new fields as 
\begin{align}
\partial^\beta F_{\alpha \beta}\,=\frac{1}{\kappa_1}j_{e\alpha}. \label{eq26} \\
\partial^\beta f_{\alpha \beta}\,=\frac{1}{\kappa_2}j_{g\alpha}. \label{eq27}
\end{align}
\subsection{Maxwell's Fields from Charge Degree of Freedom}
For classical fields, the 4-charge-current density in eq. \eqref{eq9} is represented by 
\begin{equation}\label{eq28}
j_e^\alpha = (c\rho_e,\, \mathbf{j}_e), \quad \qquad   j_{e\alpha} = (c\rho_e,\, -\mathbf{j}_e)
\end{equation}
where $\mathbf{j}_e = \rho_e\mathbf{v}$, with $\rho_e = $ electric charge density.
For static charge distributions, the current density $j_{e\alpha} = j_{e0} = c\rho_e$; it produces a time-independent - static - field, given by \eqref{eq26}:
\begin{align}
\cancelto{0}{\frac{1}{c}\frac{\partial F_{00}}{\partial t}} -\frac{\partial F_{01}}{\partial x} - \frac{\partial F_{02}}{\partial y} - \frac{\partial F_{03}}{\partial z} = \frac{\rho_ec}{\kappa_1}  \nonumber \\
\mbox{or} \quad \frac{\partial (cF_{01})}{\partial x} + \frac{\partial (cF_{02})}{\partial y} + \frac{\partial (cF_{03})}{\partial z} = \,-\, \frac{\rho_ec^2}{\kappa_1}. \label{eq29}
\end{align}
Eq. (\ref{eq29}) gives us Coulomb field ($\mathbf{E}$) as expressed in the Gauss's law of electrostatics, viz., 
\begin{equation}\label{eq30}
\mathbf{\nabla}\cdot\mathbf{E} = \frac{\partial E_{x}}{\partial x}+\frac{\partial E_{y}}{\partial y} + \frac{\partial E_{z}}{\partial z} =\frac{\rho_e}{\epsilon_0}  \qquad \mbox{(in SI units)}
\end{equation}
($\epsilon_0 = $ electric permittivity of vacuum), if we make the following identifications:
\begin{equation}\label{eq31}
F_{01} = \frac{E_x}{c},\,\,F_{02} = \frac{E_y}{c},\,\,F_{03} = \frac{E_z}{c}\,\,\&\,\,\kappa_1 = -\epsilon_0c^2.
\end{equation}
With the above value of $\kappa_1$, fixed by Coulomb's law not by us, eqs. (\ref{eq21}) and (\ref{eq26}) become  
\begin{align}
\mathcal{L}_e\,=\,-\,\frac{\epsilon_0c^2}{4}F^{\mu \nu}F_{\mu \nu}\,-\,j_e^\mu A_{e\mu}, \label{eq32} \\
\partial^\beta F_{\alpha \beta}\,=\,-\,\frac{1}{\epsilon_0c^2}j_{e\alpha}\,=\,-\,\mu_0j_{e\alpha}. \label{eq33}
\end{align}
From the anti-symmetry property of $F_{\alpha \beta}$ ($F_{\alpha \beta}\,=\,-\,F_{\beta \alpha}$), it follows form the results (\ref{eq31}) that
\begin{equation}\label{eq34}
F_{10} = - \frac{E_x}{c},\, F_{20}= - \frac{E_y}{c},\,F_{30} = - \frac{E_z}{c}\,\, \&\,F_{\alpha \alpha} = 0. 
\end{equation}
The other elements of $F_{\alpha \beta}$ can be obtained as follows. For $\alpha = 1$, i.e.  $j_{e1} = - j_{ex}$, eq.(\ref{eq33}) gives us 
\begin{align}\label{eq35}
   - \mu_0j_{e1} &= \mu_0j_{ex} \nonumber \\
   &= \partial^0F_{10} +\cancelto{0}{\partial ^1F_{11}}+\partial^2F_{12} + \partial ^3F_{13}\nonumber \\ 
                      &= -\frac{1}{c^2}\frac{\partial E_x}{\partial t} -\frac{\partial F_{12}}{\partial y} - \frac{\partial F_{13}}{\partial z} \nonumber \\
                      &= {\begin{cases}
                       -\frac{1}{c^2}\frac{\partial E_x}{\partial t} + \left(\mathbf{\nabla}\times\mathbf{B}\right)_x & \text{(For SME)}   \\
                      -\frac{1}{c^2}\frac{\partial E_x}{\partial t} - \left(\mathbf{\nabla}\times\mathbf{B}\right)_x & \text{(For NME)}
                      \end{cases}} 
\end{align}
where $F_{12} = - B_z$ and $F_{13} = B_y$ for the standard Maxwell's Equations (SME); $F_{12} = B_z$ and $F_{13} = - B_y$ for a possible form of New Maxwell's Equations (NME). This way, we determined all the elements of 
the anti-symmetric `field strength tensor' $F_{\alpha \beta}$:  
\begin{equation}\label{eq36}
F_{\alpha \beta}\,=\,\begin{cases}
\underbrace{\begin{pmatrix}
  0      &    E_x/c  &    E_y/c   &   E_z/c \\
-E_x/c   &    0      &   -B_z     &   B_y  \\ 
-E_y/c   &    B_z    &    0       &  -B_x  \\ 
-E_z/c   &   -B_y    &    B_x     &   0       
\end{pmatrix}}_{\text{For SME}}
   \\ 
\underbrace{\begin{pmatrix}
  0      &    E_x/c  &    E_y/c   &   E_z/c \\
-E_x/c   &    0      &    B_z     &  -B_y  \\ 
-E_y/c   &   -B_z    &    0       &   B_x  \\ 
-E_z/c   &    B_y    &   -B_x     &   0       
\end{pmatrix}}_{\text{For NME}}
 \end{cases}
\end{equation}
and the Amp\`{e}re-Maxwell law of SME and NME:
\begin{equation}\label{eq37}
\mathbf{\nabla}\times\mathbf{B}\,=\,\begin{cases}
\,+\,\mu_0\mathbf{j}_e\,+\,\frac{1}{c^2}\frac{\partial \mathbf{E}}{\partial t} & \quad \text{(For SME)} \\ \\
 \,-\,\mu_0\mathbf{j}_e\,-\,\frac{1}{c^2}\frac{\partial \mathbf{E}}{\partial t} & \quad \text{(For NME)} 
 \end{cases}        
\end{equation}
where the magnetic field, \textbf{B} is generated by charge current and time-varying electric field $\mathbf{E}$.\\
For reference, we note the field strength tensor with two contravariant indices:
\begin{equation}\label{eq38}
F^{\alpha \beta} =\eta^{\alpha \gamma}F_{\gamma \delta}\eta^{\delta \beta} = \begin{cases}
\underbrace{\begin{pmatrix}
  0             &   -\frac{E_x}{c}  &   -\frac{E_y}{c}   &  -\frac{E_z}{c} \\
\frac{E_x}{c}   &    0              &   -B_z             &   B_y  \\ 
\frac{E_y}{c}   &    B_z            &    0               &  -B_x  \\ 
\frac{E_z}{c}   &   -B_y            &    B_x             &   0
\end{pmatrix}}_{\text{For SME}}
 \\ \\ 
\underbrace{\begin{pmatrix}
0    &   -\frac{E_x}{c}  &   -\frac{E_y}{c}   &  -\frac{E_z}{c} \\
\frac{E_x}{c} &    0    &   B_z   &  -B_y  \\ 
\frac{E_y}{c}   &   -B_z  &    0    &  B_x  \\ 
\frac{E_z}{c}   &    B_y  &   -B_x  &   0
\end{pmatrix}}_{\text{For NME}}
\end{cases} 
\end{equation}
From eq. (\ref{eq33}) and the anti-symmetry property of $F^{\alpha \beta}$, it follows that $j_e^\alpha$ is divergence-less:
\begin{equation}\label{eq39}
\partial_\alpha j_e^\alpha\,=\,0\,=\,\,\frac{1}{c}\frac{\partial (\rho_ec)}{\partial t} + \mathbf{\nabla}\cdot\mathbf{j_e} = \mathbf{\nabla}\cdot\mathbf{j_e} \,+\,\frac{\partial \rho_e}{\partial t}.
\end{equation}
This is the {\it continuity equation} expressing the local conservation of electric charge. \\
Equation (\ref{eq33}) gives us two in-homogeneous equations of SME and NME. The very definition of 
$F_{\alpha \beta}$ in eq. (\ref{eq18}), automatically guarantees us the Bianchi identity:
\begin{equation}\label{eq40}
\partial_\alpha F_{\beta \gamma} +\partial_\beta F_{\gamma \delta} + \partial_\gamma F_{\alpha \beta} = 0,
\end{equation}
(where $\alpha , \beta, \gamma $ are any three of the integers $0, 1, 2, 3$), from which two homogeneous equations emerge naturally: 
\begin{eqnarray}
\mathbf{\nabla}\cdot \mathbf{B}\,=\,0  \quad \text{(For both SME and NME)} \label{eq41} \\ 
\mathbf{\nabla}\times \mathbf{E}= \begin{cases} 
\,-\,\frac{\partial \mathbf{B}}{\partial t} & \quad \text{(For SME)} \\ \\
\,+\,\frac{\partial \mathbf{B}}{\partial t} & \quad \text{(For NME)}
\end{cases} \label{eq42}
\end{eqnarray}

The Bianchi identity (\ref{eq40}) may concisely be expressed by the zero divergence of a dual field-strength tensor $\mathscr{F}_e^{\alpha \beta}$, viz., 
\begin{equation}\label{eq43}
\partial _\alpha \mathscr{F}_e^{\alpha \beta}\,=\,0, \quad \mbox{where $\mathscr{F}_e^{\alpha \beta}$ is defined by }
\end{equation} 
\begin{equation}\label{eq44}
\mathscr{F}_e^{\alpha \beta}\,=\,\frac{1}{2}\epsilon^{\alpha \beta \gamma \delta}F_{\gamma \delta}\,=\,\underbrace{\begin{pmatrix}
0     &   -B_x    &    -B_y    &   -B_z \\
B_x   &    0      &    \frac{E_z}{c}   &   -\frac{E_y}{c}  \\ 
B_y   &   -\frac{E_z}{c}  &    0       &   \frac{E_x}{c}   \\ 
B_z   &    \frac{E_y}{c}  &   -\frac{E_x}{c}   &     0
\end{pmatrix}}_{\text{For SME}}
\end{equation}
and the totally anti-symmetric fourth rank tensor $\epsilon^{\alpha \beta \gamma \delta}$ (called Levi-Civita Tensor) is defined by 
\begin{equation}\label{eq45}
\epsilon^{\alpha \beta \gamma \delta} =
\begin{cases}
{+1}   & \quad \text{for}\, \alpha = 0, \beta = 1, \gamma = 2, \delta = 3, and \\
{}     & \quad \text{any even permutation}          \\
{-1}   & \quad \text{for any odd permutation}  \\
{}{0}    & \quad \text{if any two indices are equal}. 
\end{cases}
\end{equation} 
The dual field-strength tensor $\mathscr{F}_e^{\alpha \beta}$ for the NME can be obtained from eq. (\ref{eq44}) by substitution $\mathbf{B} \rightarrow -\mathbf{B}$, with $\mathbf{E}$ remaining the same.\\
Eq. (\ref{eq41}) suggests that $\mathbf{B}$ can be defined as the curl of a vector function 
$\mathbf{A_e}$ (say). If we define 
\begin{equation}\label{eq46}
\mathbf{B}\,=\,\begin{cases} 
+\mathbf{\nabla}\times \mathbf{A}_e & \quad   \text{(For SME)} \\ 
-\mathbf{\nabla}\times \mathbf{A}_e & \quad   \text{(For NME)} 
\end{cases}
\end{equation}
then using these definitions in (\ref{eq42}), we find 
\begin{equation}\label{eq47}
\mathbf{\nabla}\times \left(\mathbf{E} + \frac{\partial \mathbf{A}_e}{\partial t}\right)\,=\,\mathbf{0} \quad \text{(For SME and NME)},
\end{equation}
which is equivalent to say that the vector quantity inside the parentheses of eq. (\ref{eq47}) can be written as the gradient of a scalar potential, $A_{e0}$: 
\begin{equation}\label{eq48}
\mathbf{E}\,=\,-\,\mathbf{\nabla}A_{e0} \,-\,\frac{\partial \mathbf{A}_e}{\partial t} \quad \text{(For SME and NME)}.
\end{equation}
In relativistic notation, eqs. (\ref{eq46}) and (\ref{eq48}) become 
\begin{equation}\label{eq49}
F^{\alpha \beta}\,=\,\partial^\alpha A_e^\beta \,-\,\partial^\beta A_e^\alpha,
\end{equation}
(as they must, because of their common origin) where 
\begin{equation}\label{eq50}
A_e^\alpha  = (A_{e0},\,\mathbf{A}_e) = (\phi_{e}/c,\,\mathbf{A}_e).
\end{equation}
In terms of this 4-potential, the in-homogeneous eqs. (\ref{eq33}) of SME and NME read:
\begin{equation}\label{eq51}
\partial_\beta \partial ^\beta A_e^\alpha\,-\,\partial^\alpha(\partial_\beta A_e^\beta)\,=\,\mu_0j_e^\alpha.
\end{equation}
Under the Lorenz condition,
\begin{equation} \label{eq52}
\partial_\beta A_e^\beta =\,0,
\end{equation}
the in-homogeneous equations (\ref{eq51}) simplify to the following equations:
\begin{equation}\label{eq53}
\partial_\beta \partial ^\beta {A}_e^\alpha\,=\,\Box {A}_e^\alpha\,=\,\mu_0j_e^\alpha \qquad\text{(For SME \& NME)}.
\end{equation}
The relativistic Lagrangian (not Lagrangian density) for a single particle of proper mass $m_0$ and electric charge $q$ moving in the external field of SME and NME, is written as 
\begin{equation}\label{eq54}
L_e = -\left[m_0\sqrt{\eta^{\alpha \beta}\frac{dx_\alpha}{d\tau}\frac{dx_\beta}{d\tau}}\,+\,q\frac{dx_\alpha}{d\tau}
A_e^\alpha (x)\right]
\end{equation}
Using the Lagrangian (\ref{eq54}) in Euler-Lagrange equations, one obtains the co-variant equation of motion of a charged particle in electromagnetic field:  
\begin{align} \label{eq55}
(a)\, \frac{d\dot{x}^\alpha}{d\tau} = \frac{q}{m_0}F^{\alpha \beta}u_\beta, \quad &
(b)\, \frac{dp^\alpha}{d\tau} = \frac{q}{m_0}F^{\alpha \beta}p_\beta.
\end{align}
In three dimensional form the equations of motion (\ref{eq55}b), take the following forms: \\
\begin{equation}\label{eq56}
\frac{d\mathbf{p}}{dt}\,=\,\begin{cases} 
q\left[\mathbf{E}\,+\,\mathbf{u}\times \mathbf{B}\right] & \quad   \text{(For SME)} \\ 
q\left[\mathbf{E}\,-\,\mathbf{u}\times \mathbf{B}\right] & \quad   \text{(For NME)}  
\end{cases}
\end{equation}
\begin{equation}\label{eq57}
\frac{dE}{dt}\,=\,q\mathbf{u}\cdot \mathbf{E}  \quad   \text{(For both SME and NME)} 
\end{equation}
\subsection{Maxwell-like Fields from Mass Degree of Freedom}
For classical fields the 4-current mass density or 4-momentum density in (\ref{eq10}) is represented by 
\begin{equation}\label{eq58}
j_g^\alpha = (c\rho_0,\, \mathbf{j}_g), \quad \qquad   j_{g\alpha} = (c\rho_0,\, -\mathbf{j}_g)
\end{equation}
where $\mathbf{j}_g = \rho_0\mathbf{v}$, with $\rho_0 = $ proper mass density.
For static mass distributions, the current density $j_{g\alpha} = j_{g0} = c\rho_0$. It produces a time-independent - static - field, given by eq. (\ref{eq27}). By establishing its correspondence with Newtonian gravitostatic field $\mathbf{g}$ dictated  by \,$\mathbf{\nabla}\cdot\mathbf{g} = -4\pi G\rho_0$,\, as was done for the Coulomb field in the previous section, we obtain:
\begin{equation}\label{eq59}
\kappa_2 = \frac{c^2}{4 \pi G},\, $G$\, \mbox{is Newton's gravitational constant}. 
\end{equation}
With this value of $\kappa_2$ (fixed by Newton's law, \,$\mathbf{\nabla}\cdot\mathbf{g} = -4\pi G\rho_0$,\, not by us, just as the value of $\kappa_1$ was fixed by Coulomb's law in eq. (\ref{eq30})), eqs. (\ref{eq22}) and (\ref{eq27}) turned out as 
\begin{align}
\mathcal{L}_g = \frac{c^2}{16\pi G }f^{\mu \nu}f_{\mu \nu} - j_g^\mu A_{g\mu} 
             = \frac{\epsilon_{0g}c^2}{4 }f^{\mu \nu}f_{\mu \nu} - j_g^\mu A_{g\mu} \label{eq60} \\  
\partial^\beta f_{\alpha \beta}\,=\,\frac{4\pi G}{c^2}j_{g\alpha} = \,\mu_{0g}j_{g\alpha}.              \label{eq61}
\end{align}
\noindent
where we have introduced two new constants $\epsilon_{0g}$ and $\mu_{0g}$ such that  
\begin{equation}\label{eq62}
\epsilon_{0g}\,=\,\frac{1}{4\pi G} \quad \text{and}\quad  \mu_{0g}\,=\frac{4\pi G}{c^2} \implies c \,= \,\frac{1}{\sqrt{\epsilon_{0g}\mu_{0g} }}, 
\end{equation}
in complete analogy with the electromagnetic case where $c = (\epsilon_{0}\mu_{0})^{-1/2}$. Therefore $\epsilon_{0g}$ may be called the gravitic or gravito-electric permittivity of free space and $\epsilon_{0g}$ may be called the gravito-magnetic permeability of free space. Now following the methods adopted in the previous section for discovering electromagnetic theory, we get the following results for gravito-electromagnetic (GEM) theory or what we call Heaviside-Maxwellian Gravity. The Bianchi identity for HMG:
\begin{equation}\label{eq63}
\partial_\alpha f_{\beta \gamma} +\partial_\beta f_{\gamma \delta} + \partial_\gamma f_{\alpha \beta} = 0.
\end{equation}
The gravitational analogues of eqs. (\ref{eq54})-(\ref{eq55}) are 
\begin{equation}\label{eq64}
L_g = -\left[m_0\sqrt{\eta^{\alpha \beta}\frac{dx_\alpha}{d\tau}\frac{dx_\beta}{d\tau}}\,+\,m_0\frac{dx_\alpha}{d\tau}
A_g^\alpha (x)\right]
\end{equation}
\begin{align} \label{eq65}
(a)\, \frac{d\dot{u}^\alpha}{d\tau} = f^{\alpha \beta}u_\beta, \quad &
(b)\, \frac{dp^\alpha}{d\tau} = f^{\alpha \beta}p_\beta.
\end{align}
 The anti-symmetric `field strength tensor' $f_{\alpha \beta}$ of what we call Maxwellain Gravity (MG) and Heaviside Gravity (HG):  
\begin{equation}\label{eq66}
f_{\alpha \beta}\,=\,\begin{cases}
\begin{pmatrix}
  0      &    g_x/c  &    g_y/c   &   g_z/c \\
-g_x/c   &    0      &   -b_z     &   b_y  \\ 
-g_y/c   &    b_z    &    0       &  -b_x  \\ 
-g_z/c   &   -b_y    &    b_x     &   0       
\end{pmatrix}
& \quad \text{(For MG)}   \\ \\
\begin{pmatrix}
  0      &    g_x/c  &    g_y/c   &   g_z/c \\
-g_x/c   &    0      &    b_z     &  -b_y  \\ 
-g_y/c   &   -b_z    &    0       &   b_x  \\ 
-g_z/c   &    b_y    &   -b_x     &   0       
\end{pmatrix}
& \quad \text{(For HG)} 
\end{cases}
\end{equation}
and the gravito-Amp\`{e}re-Maxwell law of MG and HG: 
\begin{equation}\label{eq67}
\mathbf{\nabla}\times\mathbf{b}\,=\,\begin{cases}
\,-\,\frac{4\pi G}{c^2}\mathbf{j}_g\,+\,\frac{1}{c^2}\frac{\partial \mathbf{g}}{\partial t} & \quad \text{(For MG)} \\ \\
 \,+\,\frac{4\pi G}{c^2}\mathbf{j}_g\,-\,\frac{1}{c^2}\frac{\partial \mathbf{g}}{\partial t} & \quad \text{(For HG)} 
 \end{cases}        
\end{equation}
where $\mathbf{b}$ is named as gravitomagnetic field, which is generated by gravitational charge (or mass) current and time-varying gravitational or gravitoelectric field $\mathbf{g}$. The field strength tensor $f^{\alpha \beta}$ is obtained as:
\begin{equation}\label{eq68}
f^{\alpha \beta} =\eta^{\alpha \gamma}f_{\gamma \delta}\eta^{\delta \beta} = \begin{cases}
\underbrace{\begin{pmatrix}
  0             &   -\frac{g_x}{c}  &   -\frac{g_y}{c}   &  -\frac{g_z}{c} \\
\frac{g_x}{c}   &    0              &   -b_z             &   b_y  \\ 
\frac{g_y}{c}   &    b_z            &    0               &  -b_x  \\ 
\frac{g_z}{c}   &   -b_y            &    b_x             &   0
\end{pmatrix}}_\text{(For MG)}
 \\ \\ 
\underbrace{\begin{pmatrix}
0    &   -\frac{g_x}{c}  &   -\frac{g_y}{c}   &  -\frac{g_z}{c} \\
\frac{g_x}{c} &    0    &   b_z   &  -b_y  \\ 
\frac{g_y}{c}   &   -b_z  &    0    &  b_x  \\ 
\frac{g_z}{c}   &    b_y  &   -b_x  &   0
\end{pmatrix}}_\text{(For HG)}
\end{cases} 
\end{equation}
The two homogeneous equations follow from the Bianchi identity eq. (\ref{eq63}) as: 
\begin{equation}\label{eq69}
\mathbf{\nabla}\cdot \mathbf{b}\,=\,0  \quad \text{(For both MG and HG)}\\
\end{equation}
\begin{equation}\label{eq70}
\mathbf{\nabla}\times \mathbf{g}= \begin{cases} 
\,-\,\frac{\partial \mathbf{b}}{\partial t} & \quad \text{(For MG)} \\ 
\,+\,\frac{\partial \mathbf{b}}{\partial t} & \quad \text{(For HG)}
\end{cases} 
\end{equation}
The eq. (\ref{eq70}) represents the gravito-Faraday's law for MG and HG.
Eq. (\ref{eq69}) suggests that $\mathbf{b}$ can be defined as the curl of a vector function 
$\mathbf{A}_g$ (say). If we define 
\begin{equation}\label{eq71}
\mathbf{b}\,=\,\begin{cases} 
+\mathbf{\nabla}\times \mathbf{A}_g & \quad   \text{(For MG)} \\ 
-\mathbf{\nabla}\times \mathbf{A}_g & \quad   \text{(For HG)} 
\end{cases}
\end{equation}
then using these definitions in eq. (\ref{eq70}), we get  
\begin{equation}\label{eq72}
\mathbf{\nabla}\times \left(\mathbf{g} + \frac{\partial \mathbf{A}_g}{\partial t}\right)\,=\,\mathbf{0} \quad
 \text{(For both MG and HG)}.
\end{equation}
So the vector quantity inside the parentheses of eq. (\ref{eq72}) is written as the gradient of a scalar potential, $A_{g0}$: 
\begin{equation}\label{eq73}
\mathbf{g}\,=\,-\,\mathbf{\nabla}A_{g0} \,-\,\frac{\partial \mathbf{A}_g}{\partial t} \quad \text{(For both MG and HG)}.
\end{equation}
In relativistic notation, eqs. (\ref{eq71}) and (\ref{eq73}) become 
\begin{align}
f^{\alpha \beta}\,=\,\partial^\alpha A_g^\beta \,-\,\partial^\beta A_g^\alpha, \quad \mbox{where} \label{eq74}  \\
A_g^\alpha  = (A_{g0},\,\mathbf{A}_g) = (\phi_{g}/c,\,\mathbf{A}_g). \label{eq75} 
\end{align}
In terms of this 4-potential, the in-homogeneous eqs. (\ref{eq61}) of MG and HG read:
\begin{equation}\label{eq76}
\partial_\beta \partial ^\beta A_g^\alpha\,-\,\partial^\alpha(\partial_\beta A_g^\beta)\,=\,-\,
\frac{4\pi G}{c^2}j_g^\alpha \,=\,-\,\mu_{0g}j_g^\alpha.
\end{equation}
Under gravito-Lorenz condition,
\begin{equation}\label{eq77}
\partial_\beta A_g^\beta =\,0,
\end{equation}
the in-homogeneous eqs. (\ref{eq76}) simplify to the following equations:
\begin{equation}\label{eq78}
\partial_\beta \partial ^\beta {A}_g^\alpha\,=\,\Box {A}_g^\alpha\,=\,-\,\mu_{0g}j_g^\alpha
 \quad\text{(For MG \& HG).}
\end{equation}
Before concluding this section we wish to note that the proper acceleration of a particle in the fields of HMG is independent of its rest mass, $m_0$ is a natural consequence of (\ref{eq65}). This is the relativistic generalization of Galileo's law of Universality of Free Fall (UFF) - known to be true both theoretically and experimentally since Galileo's time. It states that all (non-spinning) particles of whatever rest mass, moving with same proper velocity $dx_\beta/{d\tau}$ in a given gravitational field $f^{\alpha \beta}$, experience the same proper acceleration. 
In three dimensional form the equations of motion (\ref{eq65}), take the following forms: \\
\begin{equation}\label{eq79}
\frac{d\mathbf{p}}{dt}\,=\,\begin{cases} 
m_0\left[\mathbf{g}\,+\,\mathbf{u}\times \mathbf{b}\right] & \quad   \text{(For MG)} \\ 
m_0\left[\mathbf{g}\,-\,\mathbf{u}\times \mathbf{b}\right] & \quad   \text{(For HG)}  
\end{cases}
\end{equation}
\begin{equation}\label{eq80}
\frac{dE}{dt}\,=\,m_0\mathbf{u}\cdot \mathbf{g}  \quad   \text{(For both MG and HG)}. 
\end{equation}
It is to be noted that the gravito-Lorentz force law originally speculated by Heaviside by  electromagnetic analogy was of MG-type in (\ref{eq79}). The two basic sets of Lorentz-Maxwell-like Equations (ME) of gravity producing the same physical effects are given in Table 2. They represent a single vector gravitational theory, which we call Heaviside-Maxwellian Gravity (HMG). 
\section{Discussions}
The analogies and peculiar differences between Newton's law of gravitostatics and Coulomb's law of electrostatics, noted by M. Faraday \cite{19} in 1832, have been largely investigated since the nineteenth century, focusing on the possibility that the motion of masses could produce a magnetic-like field of gravitational origin - the gravitomagnetic field. After the null experimental results on the measurement of gravitomagnetic field by M. Faraday in 1849 and then again in 1859 \cite{19}, J. C. Maxwell \cite{20} tried to formulate a field theory of gravity analogous to electromagnetic theory in 1865 but abandoned it because he was dissatisfied with his results: the potential energy of a static mass distribution always negative, but he felt this should be re-expressible as an integral over field energy density which, which being the square of the gravitational field intensity, is positive \cite{15}. We note that Maxwell did a miscalculation, if one does the actual calculation analogous to electrostatic field energy \cite{21}, a negative sign comes before the square of gravitational field intensity. Later Holzm\"{u}ller \cite{22} and and Tisserand \cite{23,24} unsuccessfully attempted to explain the advance of Mercury's perihelion through Weber's electrodynamics. In 1893, Heaviside \cite{8,9,10,11,12,13,14,15} proposed a self consistent theory of gravitomagnetism and gravitational wave (GW) by writing down a set of g-MLEs (except for a sign error in the gravito-Lorentz force law), which predict transverse gravitational waves propagating in vacuum at some finite speed $c_g$ according to Heaviside-Poynting's theorem, analogous to the electromagnetic case. To complete the dynamic picture, in a subsequent paper (Part II) \cite{9,10,11,12,13,14} Heaviside speculated a gravitational analogue of Lorentz force law, in the form that comes under g-MLEs of MG in Table 1, to calculate the effect of the $\mathbf{b}$ field (particularly due to the motion of the Sun through the cosmic aether) on Earth's orbit around the Sun. Recently Behera \cite{17} (followed Galileo-Newtonian Relativistic approach) and here we found the correct form of Heaviside's speculative gravito-Lorentz force as shown in Table 1, following two independent approaches. This correction ensures that in both HG and MG, like mass currents (parallel currents) should repel each other and unlike mass currents (anti-parallel currents) should attract each other in their gravitomagnetic interaction - opposite to the case of electromagnetism where like electric currents attract each other and unlike electric currents repel each other in their magnetic interaction. Heaviside also calculated the precession of Earth's orbit around the Sun by considering his speculative force law of MG-type in Table 1 and concluded that this effect was small enough to have gone unnoticed thus far, and therefore offered no contradiction to his hypothesis that$c_g = c$. Surprisingly, Heaviside seemed to be unaware of the long history of measurements of the precession of Mercury's orbit as noted by McDonald \cite{15}, who reported Heaviside's gravitational equations (in our present notation) as given in Table 1 under the head Maxwellian Gravity (MG) - a name coined by Behera and Naik \cite{16},\footnote{Who relying on McDonald's \cite{15} report of HG, stated that MG is same as HG. This should not be taken for granted without a proof because a sign difference in some vector quantities or equations has different physical meaning/effect.}, who obtained these equations demanding the Lorentz invariance of physical laws. It is to be noted that without the correction of Heaviside's speculative gravito-Lorentz force law the effect the gravitomagnetic field of the spinning Sun on the precession of a planet's orbit has the opposite sign to the observed effect as rigtly noted by McDonald \cite{15} and Iorio and Corda \cite{25}. Apart from Maxwell and Heaviside, prior attempts to modify Newton's theory of gravitation were made by Lorentz in 1900 \cite{26} and Poincar\`{e} \cite{27} in 1905. There was a good deal of debate concerning Lorentz-covariant theory of gravitation  in the years leading up to Einstein's publication of his work in 1915 \cite{28}. For an overview of research on gravitation from 1850 to 1915, the reader may see Roseveare \cite{29}, Renn et al. \cite{30}.  Walter \cite{31} in ref. \cite{30} discussed the Lorentz-covariant theories of gravitation where no mention of Heaviside's Gravity is seen.  However, the success of Einstein's gravitation theory, described in General Relativity (see for instance \cite{28,32,33,34,35,36,37,38,39}), led to the abandonment of these old efforts. It must be noted that Einstein was unaware of Heaviside's work on gravity, otherwise 
his confidence in the correctness of Newtonian Gravity would not have been shaken as he stated before the 1913 congress of natural scientists in Vienna \cite{40}, viz.,  
\begin{quote}
\textit{For before Maxwell, electromagnetic processes were traced back to elementary laws that were fashioned as closely as possible on the pattern of Newton's force law. According to these laws, electrical masses, magnetic masses, current elements, etc., are supposed to exert on each other actions-at-a-distance that require no time for their propagation through space. Then Hertz showed 25 years ago by means of his brilliant experimental investigation of the propagation of electrical force that electrical effects require time for their propagation. In this way he helped in the victory of Maxwell's theory, which replaced the unmediated action-at-a-distance by partial differential equations.  After the un-tenability of the theory of action at distance had thus been proved in the domain of electrodynamics, confidence in the correctness of Newton's action-at-a-distance theory of gravitation was also shaken. The conviction had to force itself through that Newton's law of gravitation does not embrace gravitational phenomena in their totality any more than Coulomb's law of electrostatics and magnetostatics embraces the totality of electromagnetic phenomena. } 
\end{quote} 
Further Heaviside's work would have played the same role on equal footing as Maxwell's electromagnetic theory did in the development of special relativity. However, after Sciama's consideration \cite{41} of MG, in 1953 to explain the origin of inertia, there have been several studies on vector gravity, see \cite{14,17,42,43,44,45,46,47,48,49,50,51,52,53,54,55} and other references therein. The g-MLEs obtained here corroborate the g-LMEs obtained by several authors using a variant of classical methods: (a) Schwinger's Galileo-Newtonian Relativistic approach to get the SMLEs \cite{17, 53}, (b) Special Relativitic approaches to gravity \cite{16,52,53,54}, (c) modification of Newton's law on the basis of the principle of causality \cite{14,49} , (d) some axiomatic methods \cite{50,51} common to electromagnetism and gravitoelectromagnetism and also (e) a specific linearization scheme of General Relativity (GR) in the weak field and slow motion approximation \cite{56}. However, in the context of GR several versions of linearized approximations exist, which are not isomorphic and predict different values of speed of gravity $c_g$ in vacuum as explicitly shown by Behera \cite{53}. This is one of the limitations of GR. MG of GR origin will be denoted as GRMG below. Out of a number of linearized versions of GR considered in \cite{53}, here we pick out only 4 versions for our discussion on the value of $c_g$ below for explicit comparison and other purpose.
\subsection{On the Speed of Gravitational Waves ($c_g$)} 
It is interesting to note that our theoretical prediction on the value of $c_g = c$ precisely agree with a remarkably precise measurement of the value of $c_g = c$ with deviations smaller than a few parts in $10^{-15}$ coming from the combination of the gravitational wave event GW170817 \cite{56}, observed by the LIGO/Virgo Collaboration, and of the gamma-ray burst GRB 170817A \cite{57}. This precise measurement of $c_g$ has dramatic consequences on the viability of several theories of gravity \cite{58,59,60,61,62,63} that have been intensively studied in the last few years because many of them generically predict $c_g\neq c$. However, here we discuss below some linearized versions of GR which predict the value of $c_g\neq c$ and also $c_g = c$. 
\subsection{GRMG of Forward, Braginsky et al. and Thorne (GRMG-FBT):}
In the weak gravity and small velocity approximations of GR, the following linear gravito-Maxwell-Lorentz equations may be obtained following Forward \cite{64}, Braginsky et al. \cite{65} and Thorne \cite{66} by neglecting the non-linear terms: 
\begin{subequations}\label{eq81}
\begin{align}
\mathbf{\nabla}\cdot\mathbf{g} = - 4\pi G \rho_0,\label{eq81a} \\
\mathbf{\nabla}\times\mathbf{H}\,=\,4\left[-\,\frac{4\pi G}{c^2}(\rho_0\mathbf{v})\,+\,\frac{1}{c^2}\frac{\partial \mathbf{g}}{\partial t}\right], \label{eq81b}\\ 
\mathbf{\nabla}\cdot\mathbf{H}\,=\,0, \label{eq81c} \\
 \mathbf{\nabla}\times\mathbf{g}\,=\,-\,\frac{\partial \mathbf{H}}{\partial t} \label{eq81d} \\  
\end{align}
\end{subequations}
\begin{equation}
m_0\frac{d\mathbf{v}}{dt}= m_0\mathbf{g}+ m_0\mathbf{v}\times\mathbf{H} \label{eq82}
\end{equation}
\noindent
where $\rho_0$ is the density of rest mass, $\mathbf{v}$ is the velocity of $\rho_0$. Thorne \cite{66} noted that the only differences from Maxwell's equations are (i) the minus signs before the source terms (terms with $\rho_0$ in \eqref{eq81a} and $(\rho_0\mathbf{v})$ in \eqref{eq81b}, which cause gravity to be attractive rather than repulsive; (ii) a factor $4$ in the strength of $\mathbf{H}$, presumably due to gravity being associated with a spin-2 field rather than spin-1; (iii) the replacement of charge density by mass density times Newton's gravitation constant $G$ and (iv) the replacement of charge current density by $G\rho_0\mathbf{v}$, where $\mathbf{v}$ is the velocity of $\rho_0$. 
In empty space ($\rho_0 = 0$), the field eqs. \eqref{eq81a}-\eqref{eq81d} reduce to the following equations
\begin{subequations}\label{eq83}
\begin{align}
\mathbf{\nabla}\cdot\mathbf{g} = 0,\label{eq83a} \\
\mathbf{\nabla}\times\mathbf{H}\,=\,\frac{4}{c^2}\frac{\partial \mathbf{g}}{\partial t}, \label{eq83b}\\ 
\mathbf{\nabla}\cdot\mathbf{H}\,=\,0, \label{eq83c} \\
 \mathbf{\nabla}\times\mathbf{g}\,=\,-\,\frac{\partial \mathbf{H}}{\partial t} \label{eq83d} 
\end{align}
\end{subequations}

Now taking the curl of eq. \eqref{eq83d} and utilizing eqs. \eqref{eq83a} and  \eqref{eq83b}, we get the wave equation for the field $\mathbf{g}$ in empty space and  taking the curl of eq. \eqref{eq83b} and utilizing eqs. \eqref{eq83c} and \eqref{eq83d}, we get the wave equation for the field $\mathbf{H}$ as 
\begin{subequations}\label{eq84}
\begin{align}
\label{eq84a}
\mathbf{\nabla}^2\mathbf{g}\,-\,\frac{4}{c^2}\frac{\partial ^2\mathbf{g}}{\partial t^2} &= \mathbf{\nabla}^2\mathbf{g}\,-\,\frac{1}{c_g^2}\frac{\partial ^2\mathbf{g}}{\partial t^2} =\mathbf{0}, \\
\label{eq84b}
\mathbf{\nabla}^2\mathbf{H}\,-\,\frac{4}{c^2}\frac{\partial ^2\mathbf{H}}{\partial t^2} &= \mathbf{\nabla}^2\mathbf{H}\,-\,\frac{1}{c_g^2}\frac{\partial ^2\mathbf{H}}{\partial t^2} =\mathbf{0},
\end{align}
\end{subequations}
where $c_g\,=\,c/2$, contrary to the recent experimental data \cite{56,57}.
\subsubsection{GRMG of Ohanian and Ruffini (GRMG-OR)}
In the Non-relativistic limit and Newtonian Gravity correspondence of GR, from Ohanian and Ruffini \cite{38} (Sec. 3.4 of \cite{38}) one gets the gravito-Maxwell-Lorentz equations as 
\begin{subequations}\label{eq85}
\begin{align}
\label{eq85a}
\mathbf{\nabla}\cdot\mathbf{g} = - 4\pi G \rho_0, \\
\label{eq85b}
\mathbf{\nabla}\times\mathbf{g}= -\frac{1}{2}\frac{\partial \mathbf{H}}{\partial t}, \\ 
\label{eq85c}
\mathbf{\nabla}\cdot\mathbf{H} = 0, \\
\label{eq85d}
\mathbf{\nabla}\times\mathbf{H} = -\frac{16\pi G}{c^2} \mathbf{j} +\frac{4}{c^2}\frac{\partial \mathbf{g}}{\partial t}, \\ 
 \label{eq85e}
m_0\frac{d\mathbf{v}}{dt} = m_0[\mathbf{g}\,+\,\mathbf{v}\times \mathbf{H}]
\end{align}
\end{subequations}

\noindent
where $\rho_0$ is the (rest) mass density, $\mathbf{j}= \rho_0\mathbf{v}$ is the momentum density and the gravitational displacement term in gravito-Amp\`{e}re-Maxwell law in eq. (\eqref{eq85d}) was recently added by Behera \cite{53} to make the 
gravito-Amp\`{e}re law of Ohanian and Ruffini self consistent with the equation of continuity of rest mass. Without this added term there can not be gravitational waves in vacuum. The wave equations for the $\mathbf{g}$ and $\mathbf{H}$ fields of GRMG-OR, in vacuum now obtainable from eqs. \eqref{eq85a}-\eqref{eq85d} yield $c_g\,=\,c/\sqrt{2}$. 
\subsubsection{GRMG of Pascual-S\`{a}nchez and Moore (GRMG-PS-M):}
In some linearized scheme of GR, Pascual-S\`{a}nchez \cite{67} obtained the following gravito-Maxwell-Lorentz equations which match with Moore's findings \cite{68}:
\begin{subequations}\label{eq86}
\begin{align}
\label{eq86a}
\mathbf{\nabla}\cdot\mathbf{g} &=-4\pi G\rho_0, \\
\label{eq86b}
\mathbf{\nabla}\times\mathbf{g}&= -\frac{\partial \mathbf{b}}{\partial t}, \\ 
\label{eq86c}
\mathbf{\nabla}\cdot\mathbf{H} = 0, \\
\label{eq86d}
\mathbf{\nabla}\times\mathbf{b} &= - \frac{4\pi G}{c^2}\mathbf{j}+\frac{1}{c^2}\frac{\partial \mathbf{g}}{\partial t}, \\ 
 \label{eq86e}
m_0\frac{d\mathbf{v}}{dt} = m_0[\mathbf{g}\,+\,4\mathbf{v}\times \mathbf{H}]
\end{align}
\end{subequations}
where $\mathbf{j}=\rho_0\mathbf{v}$. The waves equations in vacuum that emerge from eqs. \eqref{eq86a}-\eqref{eq86d} give us $c_g = c$. But note a factor of $4$ in the gravitomagnetic force term in eq.\eqref{eq86e}, which defies correspondence principle. 
\subsubsection{GRMG of Ummarino-Gallerati (GRMG-UG):}
In another linearized approximations of GR, recently Ummarino and Gallerati \cite{55} derived the following gravito-Lorentz-Maxwell equations from Einstein's GR. The gravito-Maxwell's equations are the same as those of GRMG-PS-M in eq. \eqref{eq86a}-\eqref{eq86d}, but the gravito-Lorentz force is  
\begin{equation} \label{eq87}
m_0\frac{d\mathbf{v}}{dt} = m_0\mathbf{g}+ m_0\mathbf{v}\times \mathbf{b}.
\end{equation}
\noindent
The field equations of GRMG-UG yield $c_g = c$ in vacuum. Note the absence of the factor of $4$ in grvaito-Lorentz equation. The Maxwell-Lorentz equations of GRMG-UG match with the non-relativistic limit of our findings here.  \\ 
Thus the reader can now realize that the predictions on the speed of gravity in the weak field and slow motion approximation of GR are not unique, but the value of $c_g$ is uniquely and unambiguously fixed at $c_g= c$ in the present field theoretical findings of HMG or our previous findings \cite{16,53}. It is interesting to note that the existence of gravitational waves has recently been detected \cite{69,70,71,72} and also the existence of the gravitomagnetic field generated by mass currents has been confirmed by experiments \cite{73,74,75,76,77,78,79}. These are being considered as new confirmation tests of GR \cite{69,70,71,72,73,74,75,76,79}. The explanations for experimental data on gravitational waves and the gravitomagnetic field within the framework of HMG are being explored by the authors, since the explanations for the (a) perihelion advance of Mercuty (b) gravitational bending of light and (c) the Shapiro time delay within the vector theory of gravity exist in the literature \cite{46,47,48,80}. Recently Hilborn \cite{81} following an electromagnetic analogy, calculated the wave forms of gravitational radiation emitted by orbiting binary objects that are very similar to those observed by the Laser Interferometer Gravitational-Wave Observatory (LIGO-VIRGO) gravitational wave collaboration in 2015 up to the point at which the binary merger occurs. Hilborn's calculation produces results that have the same dependence on the masses of the orbiting objects, the orbital frequency, and the mass separation as do the results from the linear version of general relativity (GR). But the polarization, angular distributions, and overall power results of Hilborn differ from those of GR. Very recently we have reported an undergraduate level explanation of the Gravity Probe B experimental results (of NASA and Stanford University) \cite{77,78,79} using the HMG \cite{82}. 
\subsection{Does GR satisfy the correspondence principle?}
By deducing Newtonian Gravity (NG) from GR, all texts books on GR teach us that GR does satisfy the correspondence principle by which a more sophisticated theory should reduce to a theory of lesser sophistication by imposing some conditions; Misner, Thorne and Wheeler \cite{32} in a boxed item (Box 17.1, page-412) of their book ``Gravitation" have put much emphasis on it by giving a host of examples. In the light of our findings on HMG here and in \cite{53} we see that GR defies the correspondence principle (cp) in its true sense: GRMG $ \cancel{\Leftrightarrow}$ SRMG $\Leftrightarrow$ N(R)MG $\Leftrightarrow$ NG, where SRMG, N(R)MG and NG stands for Special Relativistic Maxwellian Gravity, Non-relativistic or Newtonain MG and Newtonian Gravity respectively.   
\subsection{Misner, Thorne and Wheeler on HMG and Experimental Tests of HMG}
Misner,Thorne and Wheeler (MTW)\cite{32}, in their ``Exercises on flat space-rime theories of gravity", have considered a possible vector theory of gravity within the framework of special relativity. They considered a Lagrangian density of the form \eqref{eq60} and found it to be deficient in that there is no bending of light, perihelion advance of Mercury and gravitational waves carry negative energy in vector theory of gravity. As regards the classical tests of the GR, we have noted before that the explanation of these tests exist in the literature \cite{46,47,48,80}. But the issue of energy and momentum carried by gravitational waves is far from clear yet, even within the framework of GR. In the community of general relativists, there is no unanimity of opinion on the energy carried by gravitational waves. For instance, one finds references in the literature on GR which describes (not in the gravito-electromagnetic approach) the radiation from a gravitating system as carrying away energy \cite{32,83}, bringing in energy \cite{84}, carrying no energy \cite{85} or having an energy dependent on the coordinate system used \cite{85}.  However, in the gravito-electromagnetic approach to gravitational waves we briefly show that gravitational waves carry positive energy in accordance with the continuity equation or gravitational Heaviside-Poynting's theorem in spite of the fact that intrinsic energy of static gravitoelectromagnetic fields is negative. 
\subsubsection{Gravito-Maxwell Displacement Current and Continuity of Gravitoelectric Current}
Let us recall that in Maxwell's theory the displacement current is responsible for electromagnetic waves carrying energy and momentum in accordance with the continuity equation. Similar things occur in gravitoelectromagnetic theory under discussion. To see this consider the integral form of gravito-Amp\'{e}re-Maxwell Equation of MG: 
\begin{equation}
\oint \mathbf{b}\cdot d\mathbf{l} = - \mu_{0g}(I_c + I_d), \label{eq88} 
\end{equation}
where,
\begin{eqnarray}
I_d = -\epsilon_{0g}\frac{d\Psi_g}{dt},  \label{eq89} \\ 
\Psi_g = \oint \mathbf{g}\cdot d\mathbf{s} = \mbox{gravitoelectric flux}, \label{eq90}
\end{eqnarray}  
$I_c$ is the conduction current of mass and $I_d$ is the displacement current. \\
Consider a closed surface enclosing a volume. Suppose some mass is entering the volume and some mass also leaving the volume. If no mass is accumulated inside the volume, total mass going into the volume in any time is equal to the total mass leaving it during the same time. The conduction current of mass is continuous. \\
If mass is accumulated inside the volume, as in the case coalescence of two massive objects such as two neutron stars or any massive objects, this continuity breaks.  However, if we consider the conduction mass current plus the gravitational displacement current, the total current is still continuous. Any loss of conduction mass current $I_c$ appears as gravitational displacement current $I_d$. This can be shown as follows. \\
Suppose a total conduction mass current $I_1$ goes into the volume and a total conduction mass current $I_2$ goes out of it. The mass going into the volume in a time $dt$ is $I_1dt$ and that coming out is $I_2dt$. The mass accumulated inside the volume is 
\begin{equation}
dm_0^{inside} = I_1dt - I_2dt \quad \mbox{or} \quad \frac{d}{dt}(m_0^{inside}) = I_1 - I_2 \label{eq91} 
\end{equation}
From Gauss's law:
\begin{equation}
\Psi_g = \oint \mathbf{g}\cdot d\mathbf{s} = - \frac{m_0^{inside}}{\epsilon_{0g}} \label{eq92}
\end{equation}
From eqs. \eqref{eq89},\eqref{eq91} and \eqref{eq92} we get: 
\begin{equation}
I_1 - I_2 = I_d \quad 
\mbox{or} \quad I_1 = I_2 + I_d. \label{eq93}
\end{equation}

Thus total conduction current going into the volume is equal to the total current (conduction + displacement) going out of it. Note that since $\Psi_g $ is negative, $I_d$ is positive, which carries positive field momentum and energy as no actual mass is moving in such current. So gravitational collapse always leads to positive field energy and momentum coming out in the form of gravitational radiation.
\subsubsection{Gravitational Heaviside-Poynting's theorem}
The forms of the laws of conservation of energy and momentum are important results to establish for the gravitoelectromagetic field. Following the methods of electromagnetic theory, we obtained the following law of conservation of energy expressed by what we call Heaviside-Poynting's theorem as Heaviside first considered such a law (with a wrong sign for the Poynting vector) in his theory of gravity. The mathematical form of this theorem, in the form of a differential conservation law, is obtained for MG as 
\begin{eqnarray}
\frac{\partial |u|}{\partial t} + \mathbf{\nabla}\cdot \mathbf{S}= \mathbf{j}\cdot\mathbf{g} \label{eq94} \\
\mbox{where} \quad  |u| = \frac{1}{2}\epsilon_{0g}(\mathbf{g}^2 + c^2\mathbf{b}^2), \quad \mathbf{S} = \frac{1}{\mu_{0g}}(\mathbf{g}\times \mathbf{b}). \label{eq95}  
\end{eqnarray}
Note that the sign before the source term $\mathbf{j}$ is positive, whereas in the electromagnetic case the sign is negative. This is because the source terms in the field equations of MG has opposite sign to that in standard Maxwell's equations. 
The integration of $\mathbf{j}\cdot\mathbf{g}$ over a fixed volume is the total rate of doing work by the fields in that volume, which is always positive for self gravitating systems. So the right hand side of the differential energy conservation law in eq. \eqref{eq94} is positive. The vector $\mathbf{S}$, representing the energy flow, is the Heaviside-Poynting vector. The work done per unit time per unit volume by the fields 
$(\mathbf{j}\cdot\mathbf{g})$ is a conversion of gravitoelectromagnetic energy into mechanical or heat energy. Thus there is a decrease in field energy density $ = - \partial u/\partial t$. Since field energy density $u = -|u|$, we have $ - \partial u/\partial t = \partial |u|/\partial t$, which is positive. So positive energy flux of field energy must come out of systems collapsing under self gravity.   
\subsection{On the spin of graviton: spin $1$ or spin $2$?}
Following the usual procedures of electrodynamics (see, for instance \cite{86}) for obtaining the spin of photon, the spin of graviton (a quantum of gravitational wave carrying energy and momentum) in the framework of HMG can be shown to be $1$ in the unit of $\hbar$. Regarding the idea of spin-2 graviton, Wald \cite{33}(see p.76) noted that the linearized Einstein's equations in vacuum are precisely the equations written down by Fierz and Pauli \cite{87}, in 1939, to describe a massless spin-2 field propagating in flat space-time. {\it Thus, in the linear approximation, general relativity reduces to the theory of a massless spin-2 field which undergoes a non-linear self- interaction. It should be noted, however, that the notion of the mass and spin of a field require the presence of a flat back ground metric $\eta_{ab}$ which one has in the linear approximation but not in the full theory, so the statement that, in general relativity, gravity is treated as a mass-less spin-2 field is not one that can be given precise meaning outside the context of the linear approximation} \cite{33}. Even in the context of linear approximations, the original idea of spin-2 graviton gets obscured due to the several faces of non-isomorphic Gravito-Maxwell equations seen in the literature from which a unique and unambiguous prediction on the spin of graviton is difficult to get \cite{53}.
\subsection{Attraction Between Static Like Masses}
It is frequently overlooked that the interaction between two static (positive) masses, in a linear gravitational theory such as the MG or linearized vesions of GR listed earlier here, is definitely attractive. This fact was clearly understood and stated by Sciama \cite{41} and Thorne \cite{66}, who attributed this attaction to the sign before the source terms of gravito-Maxwell equations, but rarely recognized. However, to see it explicitly, let us find the static interaction between two neutral point (positive) masses at rest  within the framework of  Maxwellian Gravity, following two approaches: (1) a classical approach by Shapiro and Teukolsky \cite{88} and (B) Feynman's \cite{2} quantum field theoretical approach already described for the electrostatic case as follows. \\
\noindent
\subsubsection{Classical Approach of Shapiro and Teukolsky \cite{88}.}
For a neutral particle having gravitational charge $m_g = m_0$ at rest at the origin, the 4-current densities:
\begin{align}
j_g^0 = m_0c\delta^3(\mathbf{x}), & \qquad \mathbf{j}_g = \mathbf{0} \label{eq96} \\
\mbox{In eq. \eqref{eq77}, we put} \quad &  A_g^0 = \phi_g/c, \qquad  \mathbf{A}_g = \mathbf{0}  \label{eq97}
\end{align}
to get
\begin{equation}\label{eq98}
\mathbf{\nabla}^2\phi_g = \mu_{0g} c^2 m_0 \delta^3(\mathbf{x})\, = 4\pi G m_0 \delta^3(\mathbf{x}).
\end{equation}
This is nothing but the Poisson's equation for gravitational potential of a point mass at rest at origin. Using Green's Function, the potential at a distance $r$ for a central point particle having gravitational mass $m_0$ (i.e., the fundamental solution) is
\begin{equation}\label{eq99}
\phi_g(r) = -\frac{Gm_0}{r},
\end{equation}
which is equivalent to Newton's law of universal gravitation. The interaction energy of two point particles having gravitational charges $m_0 = M_1$ and $m_0^\prime = M_2$ separated by a distance $r$ is 
\begin{equation}\label{eq100}
U_{12}^g = M_2 \phi_g = - \frac{G M_1M_2}{r},
\end{equation}
which is {\it negative} for like gravitational charges and {\it positive} for unlike gravitational charges, if they exist. With $M_1$ at rest at the origin, the force on another stationary gravitational charge $M_2$ at a distance $r$ from origin is 
\begin{equation}\label{eq101}
\mathbf{F}_{21}^g = - M_2\mathbf{\nabla}\phi_g(r)= -\frac{GM_1M_2}{r^2}\mathbf{\hat{r}} = - \mathbf{F}_{12}^g.
\end{equation}
This force is attractive, if $M_1$ and $M_2$ are of same sign and repulsive if they are of opposite sign, unlike the case of electrical interaction between two static electric charges.  
\subsubsection{Quantum Field Theoretical Approach of Feynman \cite{2}.}
Analogous to the case of electromagnetism, the source of gravito-electromagnetism\footnote{This term is coined because of the analogy of electromagnetism with HMG.} is the the vector current $j_g^\alpha$, which is related to vector potential $A_g^\alpha$ by the relation 
\begin{eqnarray}
A_g^\alpha = \frac{4\pi G }{c^2}\frac{1}{k^2}j_g^\alpha. \label{eq102} \\ 
\mbox{In electromagnetism:}\, A_e^\alpha = - \mu_0\frac{1}{k^2}j_e^\alpha. \label{eq103}
\end{eqnarray}

\noindent
Here we have taken Fourier transforms and used the momentum-space representation. The  
 D'Alembertian operator ($\partial_\beta \partial^\beta$) in eq. \eqref{eq78} is simply $-k^2$ in momentum-space. As in electromagnetism the calculation of amplitudes in gravito-electromagnetism is made with the help of propagators connecting currents in the manner as symbolized by Feynman diagrams as that in Figure 1. The amplitudes for such processes are generally computed as a function of relativistic invariants restricting the answer as demanded by rules of momentum and energy conservation. As in electromagnetism, the guts of gravito-electromagnetism are contained in the specification of the interaction between a mass current and the field as $j_{g\alpha} A_g^\alpha$; in terms of the sources, this becomes an interaction between two currents:
 \begin{figure}
 \centering
 \includegraphics[scale=0.40]{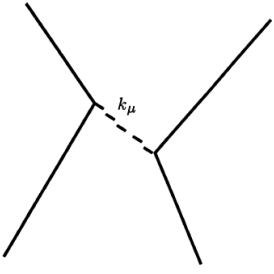}
 \caption{Feynman Diagram}
 \end{figure}
 
\begin{equation} \label{eq104}
j_{g\alpha}^\prime A_g^\alpha = \frac{4\pi G }{c^2} j_{g\alpha}^\prime \frac{1}{k^2}j_g^\alpha.
\end{equation}
In our choice of coordinates and units $k^\alpha = (\omega/c, 0, 0, \kappa)$, \, $k_\alpha = (\omega/c, 0, 0, -\kappa)$ and $A_g^\alpha$ is given by eq. \eqref{eq75}. Then the current-current interaction when the exchanged particle has a momentum $k^\alpha$ is given by the right hand side of eq. \eqref{eq104} as 
\begin{eqnarray}
\frac{4\pi G }{c^2}\frac{1}{\omega^2/c^2 - \kappa^2 }\left(j_{g0}^\prime j_g^0 -  j_{g1}^\prime j_{g1} - j_{g2}^\prime j_{g2} - j_{g3}^\prime j_{g3} \right) \nonumber \\ 
=\frac{4\pi G}{\omega^2 - c^2\kappa^2 }\left(j_{g0}^\prime j_{g0} -  j_{g1}^\prime j_{g1} - j_{g2}^\prime j_{g2} - j_{g3}^\prime j_{g3} \right) \label{eq105}
\end{eqnarray}

The conservation of proper mass, which states that the four divergence of proper mass current is zero, in momentum-space becomes simply the restriction 
\begin{equation}
k^\alpha j_{g\alpha} = 0. \label{eq106}
\end{equation}  
In the coordinate system we have chosen, this restriction connects the third and the zeroth component of the currents by 
\begin{equation}\label{eq107}
 \frac{\omega}{c}j_{g0} - \kappa j_{g3} = 0, \quad  \mbox{or} \quad  j_{g3} = \frac{\omega}{\kappa c}j_{g0}. 
\end{equation} 
If we insert this expression for $j_{g3}$ into eq. \eqref{eq105}, we get
 \begin{equation}\label{eq108}
 j_{g\alpha}^\prime A_g^\alpha = - \frac{4\pi G }{\kappa^2c^2} j_{g0}^\prime j_{g0} - \frac{4\pi G}{\omega^2 - c^2\kappa^2 }\left(j_{g1}^\prime j_{g1} + j_{g2}^\prime j_{g2} \right) 
\end{equation}  
Now we can give interpretation to the two terms in eq. \eqref{eq108}. The zeroth component of the current is simply the mass density; in the situation where we have stationary masses, it is the only on-zero component of current. The first term is independent of frequency; when we take the inverse Fourier  transform  to convert this to a space-interaction, we find that it represents an instantaneously acting Newton potential. 
 \begin{equation}\label{eq109}
 (F.T.)^{-1} \left[ - \frac{4\pi G }{\kappa^2c^2} j_{g0}^\prime j_{g0}\right] = - \frac{Gm_0^\prime m_{0}}{r}\delta(t - t^\prime).  
\end{equation}  
This is always the leading term in the limit of small velocities. The term appears instantaneous, but this is only due to the separation we have made into two terms is not manifestly co-variant. The total interaction is really an invariant quantity; the second term represents corrections to the instantaneous Newtonian interaction. The force in eq. \eqref{eq109} is attractive, if $m_{0}$ and $m_0^\prime$ are of the same sign and repulsive if they are of the opposite sign - the reverse case of electrical interaction between two static electric charges. Besides the above two different approaches, one may adopt Zee's \cite{5} path-integral approach to get at the same conclusion if one uses our equations  \eqref{eq60} and \eqref{eq78}. 
\section{Conclusion}
 We have arrived at Maxwell-Lorentz electrodynamics using the principle of local phase invariance applied to the free Dirac Lagrangian by considering the phase associated with the electric charge of a Dirac particle and Coulomb's law corresponding to the static part of the Dirac particle charge density in its classical limit. Free Dirac Lagrangian density eq. \eqref{eq1} when combined with eq. \eqref{eq32} with $j_e^\mu = qc(\overline{\psi}\gamma^\mu\psi)$ as in eq. \eqref{eq9} one obtains the Lagrangian density for quantum electrodynamics - charged Dirac fields (electrons and positrons) interacting with Maxwell's fields (photons). This is truly a breathtaking accomplishment as Griffiths \cite{85} states it; because the requirement of local phase invariance associated with the charge of the Dirac particle, applied to the free Dirac Lagrangian density, generates all of electrodynamics and specifies the charge current produced by charged Dirac particles. From 1820 when Oersted discovered magnetic effects of electric current, through Faraday's discovery of electromagnetic induction in 1831 to Maxwell's synthesis of all experimental laws of electromagnetism and prediction of electromagnetic waves and their subsequent observation by Hertz in 1887, people took almost 70 years to understand the nature of classical electromagnetic phenomena. But the principle of local phase invariance led us to arrive at the Maxwell's equations almost with no time in comparison with the 70 years. This shows the predictive power of the principle of local phase invariance regarding the nature of the fields and their interactions with their sources, which we apply here to the mass degree of freedom of the same charged particle in exploring the nature of gravitational field and its interaction with its sources. Inspired by this successful story of local phase invariance and out of scientific curiosity, here in this work we applied the same principle of local phase (gauge) invariance of field theory to the Lagrangian of a free Dirac charged particle in flat space-time and explored the question, `What would result if the total phase comes from two independent contributions associated with the charge and mass degrees of freedom of charged Dirac particle?'  As a result of this study we found two independent vector fields one describing Maxwell's theory and the other is a rediscovery of Heaviside's Gravity of 1893. The important findings of this curious study include:
\begin{enumerate}
\item[(1)] a new set of Maxwell-Lorentz Equations (n-MLEs) of electromagnetism which is physically equivalent to the standard set of equations; these n-MLEs has also been found by the 1st author using Schwinger's non-relativistic formalism,
\item[(2)] a field theoretical derivation of the field equations of Heaviside's Gravity (HG) and Maxwellian Gravity (MG) as well as their respective Lorentz force laws in which we found a correction to the Heaviside's speculative gravito-Lorentz force,   
\item[(3)] our findings that HG and MG are mere two different mathematical representations of a single theory which we named as Heaviside-Maxwellian Gravitoelectromagnetism or Gravity (HMG), 
\item[(4)] the gravito-Lorentz-Maxwell's equations of MG derived here using the well established principle of local phase (or gauge) invariance of field (particularly Dirac's spinor field: a quantum field) theory perfectly match with those obtained from a variant other established methods of study or principles of classical physics: (a) Schwinger's formalism based on Galilio-Newtonian relativity if $c_g = c$, (b) special relativistic approaches of different types, (c) principles of causality, (d) some axiomatic approaches common to electromagnetism and gravitoelectromagnetism and (e) in some specific linearization method of general relativity,
\item[(5)] Galileo's Law of Universality of Free Fall is a consequence of HMG, not an initial assumption as in Einstein's General Relativity (GR), 
\item[(6)] our prediction of an unambiguous and unique value of speed of gravitational waves ($c_g = c$), which agree very well with recent experimental data, unlike the the ambiguous and non-unique value of $c_g$ obtainable from different linearized versions of GR,
\item[(7)] possible existence of spin-1 graviton, in contrast with the idea of spin-2 graviton within GR - an idea not well founded in the GR,  
\item[(8)] that the spin-1 vector gravity of HMG denomination produces attractive interaction between like static masses contrary to the prevalent view of the field theorists, 
\item[(9)] a brief discussion on the issue of negative/positive energy of gravitational waves both in HMG and GR and two theoretical demonstrations that gravitational waves emanating from the collapsing process of self gravitating systems carry positive energy and momentum in the spirit of Maxwell's electromagnetic theory despite the fact that the intrinsic energy of static gravitoelectromagnetic fields is negative as dictated by Newton's gravitational law and its time-dependent extensions to HMG, 
\item[(10)] a gravitational correction to the standard Lagrangian of  electrodynamics, which, for a neutral massive Dirac particle, reduces to the Lagrangian for gravitodynamics and
\item[(11)] our mention of the works of some other researchers which correctly explain some crucial test of GR, viz., (a) non-Newtonain perihelion advance of planetary orbits including Mercury,(b) gravitational bending of light, (c) Shapiro time delay and (d) gravitational wave forms of recently detected gravitational waves, in a non-GR approach but using some aspects of HMG. 
\end{enumerate}
Being simple, self consistent and well founded, HMG may deserve certain attention of the researchers interested in probing the classical and quantum gravitodynamics of moving bodies/particles in the presence and absence of electromagnetic or other interactions having energy-momentum 4-vector, which couples to the 4-vector potential of HMG. 
This work, while corroborating previous works of several researchers, presents theoretical results of immediate impact that hold potential to initiate new avenues of research on quantum gravity. It also provides compelling new preliminary results on controversial, long-standing questions of localization and transfer of gravitational field energy in the form of gravitational waves and presents a concise conceptual advance on the long neglected and often rejected theory of Heaviside-Maxwellian Gravity.


%
%


\begin{thebibliography}{}
\bibitem{13} An unedited copy of the original Heaviside's article, except that some formulas and all vector equations have been converted to modern notation, is reproduced in \cite{14} below, p. 189-202.
\end{thebibliography}


\end{document}